\documentclass[twoside,leqno,twocolumn]{article}
\pdfoutput=1

\usepackage{siamproceedings}
\usepackage[T1]{fontenc}
\usepackage{lmodern}
\usepackage{graphicx}
\usepackage{url}
\usepackage{booktabs}
\usepackage[utf8]{inputenc}
\usepackage[nopatch=footnote]{microtype}
\usepackage{amsfonts}
\usepackage{amssymb}
\usepackage{amsmath}
\usepackage[font=small,labelfont=bf]{caption}
\usepackage{subcaption}
\usepackage{setspace}
\usepackage{etoolbox}

\hypersetup{
  hidelinks,
  bookmarksopen,
  bookmarksnumbered,
  colorlinks,
  urlcolor = blue,
  linkcolor = blue,
  citecolor = blue
}

\allowdisplaybreaks

\makeatletter
\renewcommand\paragraph{\@startsection{paragraph}{4}{.25in}%
  {1.0ex\@plus .1ex \@minus .1ex}%
  {-.5em plus -.1em}%
  {\reset@font\normalsize\bfseries}}
\makeatother

\patchcmd{\thebibliography}{\settowidth}%
  {\setlength{\itemsep}{0pt plus 0.30ex}\settowidth}{}{}

\newcommand{\bigO}{\mathcal{O}}
\newcommand{\dd}{\mathinner{.\,.}}

\newcommand{\Z}{\mathbb{Z}}

\newcommand{\Zp}{\Z_{>0}}

\newcommand{\Text}{T}

\newcommand{\Textlen}{n}
\newcommand{\AlphabetSize}{\sigma}
\newcommand{\IntegerAlphabet}{[0 \dd \AlphabetSize)}

\newcommand{\emptystring}{\varepsilon}
\newcommand{\LCE}[3]{\mathrm{LCE}_{#1}(#2,#3)}

\newcommand{\SubstringComplexitySym}{\delta}
\newcommand{\LZSizeSym}{z}
\newcommand{\LZNonOvSizeSym}{z_{\rm no}}

\newcommand{\SubstringComplexity}[1]{\SubstringComplexitySym(#1)}
\newcommand{\LZSize}[1]{\LZSizeSym(#1)}
\newcommand{\LZNonOvSize}[1]{\LZNonOvSizeSym(#1)}

\newcommand{\Rhs}[2]{\mathrm{rhs}_{#1}(#2)}
\newcommand{\Exp}[2]{\mathrm{exp}_{#1}(#2)}
\newcommand{\Lang}[1]{L(#1)}

\title{Engineering Fast and Space-Efficient Recompression from SLP-Compressed
   Text\thanks{The full version of this paper is available at
   \protect\url{https://arxiv.org/abs/2506.12011}.
   This work has been partially funded by the NSF CAREER Award
   2337891 and the Simons Foundation Junior Faculty Fellowship.}}

\author{
  Ankith Reddy Adudodla\thanks{
    Department of Computer Science,
    Stony Brook University,
    Stony Brook, NY, USA,
    \texttt{aadudodla@cs.stonybrook.edu}.
  }
  \and
  Dominik Kempa\thanks{
    Department of Computer Science,
    Stony Brook University,
    Stony Brook, NY, USA,
    \texttt{kempa@cs.stonybrook.edu}.
  }
}

\date{}

\begin{document}

\maketitle

\begin{abstract}
  Compressed indexing enables powerful queries over massive and
  repetitive textual datasets using space proportional to the
  compressed input. While theoretical advances have led to highly
  efficient index structures, their practical construction remains a
  bottleneck, especially for complex components like recompression
  RLSLP --- a grammar-based representation crucial for building powerful
  text indexes that support widely used suffix and LCP array queries.

  In this work, we present the first implementation of recompression
  RLSLP construction that runs in compressed time, operating on an
  LZ77-like approximation of the input. Compared to state-of-the-art
  uncompressed-time methods, our approach achieves up to
  $46\times$ speedup and $17\times$ lower RAM usage on large,
  repetitive inputs. These gains unlock scalability to larger datasets
  and affirm compressed computation as a practical path forward for
  fast index construction.
\end{abstract}

\section{Introduction}\label{sec:intro}

Data compression is a classic area in computer science, where the goal
is to reduce the size of a given file for transfer (e.g., over a slow
network) or for storage. In recent years, however, compression has
been applied in a new way. The field of \emph{compressed indexing}
aims to store any string $\Text \in \Sigma^{\Textlen}$ of $\Textlen$
symbols over an alphabet $\Sigma$ in space proportional to the size of
$\Text$ in compressed form, while simultaneously supporting queries
over the (uncompressed) text $\Text$.

Such functionality is a fundamental part of the toolbox in
applications where massive textual datasets exhibit significant
redundancy, including storage and retrieval of source code
repositories (such as GitHub)~\cite{NavarroIndexes,NavarroMeasures},
versioned text documents (such as Wikipedia), and computational
biology, where massive genomic datasets---due to projects like the
100,000 Genomes Project~\cite{100k} or the ongoing 1+ Million Human
Genome Initiative~\cite{mg}---produce datasets on the order of
terabytes (and are projected to reach
exabytes~\cite{estimate,stephens2015big}).

The queries currently supported by compressed text indexes include a
rich repertoire of operations, ranging from basic queries such as random
access to $\Text$~\cite{charikar,Rytter03,blocktree,BLRSRW15,attractors,KempaS22}
to more complex longest common extension (LCE)
queries~\cite{tomohiro-lce,dynstr,collapsing,DuysterK24}, and even
full suffix array and LCP array functionality~\cite{Gagie2020,collapsing}.

One of the biggest and most widely recognized challenges in the field
of compressed indexing is index
\emph{construction}~\cite{NavarroIndexes}. This is particularly
difficult for the most powerful indexes,
such as~\cite{Gagie2020,collapsing}, which are capable of supporting
the widely utilized suffix and LCP array queries~\cite{gusfield}.
Continuous progress on this problem has resulted in algorithms that
achieve improved
scalability~\cite{BoucherGKM18,KuhnleMBGLM19,OlivaGB23,%
  BingmannGK18,BingmannD0KO022,KurpiczM0S24,fsais,pSAscan,eSAIS}, but
index construction remains a major bottleneck in
practice~\cite{NavarroIndexes}.

A common feature of all the above algorithms is that they
construct the index in a single step; that is, the algorithm takes the
text $\Text \in \Sigma^{\Textlen}$ as input and, in $\bigO(\Textlen)$
time, directly outputs the index. However, in the worst case,
these methods use $\Theta(\Textlen)$ extra disk or RAM
space, even for compressible strings.

Recently, an alternative method for constructing powerful text indexes
was proposed in~\cite{resolution,collapsing}. In this method, the
index is constructed by first applying a lightweight but fast
compression that shrinks the input text close to the size of its
LZ77-compressed representation
(e.g.,~\cite{relz}). This step takes $\bigO(\Textlen)$ time and
shrinks the input text $\Text$ down from $n$ characters to, say,
$C(\Text)$ bytes, where $C(\Text) \ll \Textlen$. Following this step
is a sequence of algorithms operating on the compressed representation
of $\Text$, all taking time $\bigO(C(\Text) \log^{\bigO(1)}
\Textlen)$, i.e., roughly proportional to the size of the data in
compressed form. This new approach to index construction is still in
its early stages of adoption in practice, as each of the steps
running in compressed time is usually a complex algorithm that relies
on not-yet-implemented data structures.

Recently, however, the potential of compressed computation has started
to materialize, and one of the first algorithms running in compressed
time was implemented in~\cite{KempaL21}. More
specifically,~\cite{KempaL21} implemented an algorithm that takes the
LZ77-like compressed representation of size $C(\Text)$ of the input
text $\Text$ as input and, in $\bigO(C(\Text) \log^{\bigO(1)} n)$ time,
outputs a grammar-compressed representation
of $\Text$ (a variant of the so-called \emph{AVL grammar}~\cite{Rytter03}).
Indeed,~\cite{KempaL21} showed that this
conversion represents only a tiny fraction of the time needed for the
initial LZ77 approximation, confirming the potential of
compressed computation.

In this work, we continue the quest for achieving fast index
construction via compressed computation and address the second step
in the pipeline of index construction algorithms proposed
in~\cite{resolution,collapsing}, i.e., the conversion of the
grammar-compressed text into the so-called recompression
RLSLP~\cite{Jez15,Jez16}. Recompression RLSLP is a compressed
representation of a text $\Text$ as a context-free grammar (CFG)
producing only $\Text$ that additionally satisfies the \emph{local
  consistency} property; that is, different occurrences of the same
substring of $\Text$ are represented in a very similar way in the
grammar. Recompression RLSLP represents one of the most complex steps
in the pipeline of compressed-time index construction
from~\cite{resolution,collapsing}---it is the basis of all indexes
for the so-called ``internal pattern matching'' (see Section~5
in~\cite{resolution} and Section~5 in~\cite{collapsing}), which are
the key data structures in the construction of indexes like those
in~\cite{resolution,collapsing}.

Unlike in the case of~\cite{KempaL21} (where no prior implementation
existed for the problem addressed in the paper), construction of the
recompression RLSLP \emph{does} already have a highly engineered (and
parallelized) implementation~\cite{recompression-github}, which runs
in $\bigO(\Textlen)$ time. Given the importance of efficient
construction of recompression RLSLP in building powerful
text indexes, we thus ask:

\begin{center}
  \emph{Can we speed up the construction of recompression RLSLP by
    using compressed computation?}
\end{center}

\paragraph{Our Results}

We present the first implementation of recompression running in
compressed time. Compared to the state-of-the-art implementation of
recompression~\cite{recompression-github}, our modular pipeline, where
we first approximate the LZ77 (we developed our own prototype of the
Bentley--McIlroy LZ77 approximation~\cite{bentley1999dcc}) and then
construct recompression in compressed time, achieves the following:

\begin{itemize}

\item \emph{Drastic time reduction}: On a highly repetitive file of
  size 4\,GiB, our implementation is $46\times$ faster than the
  state-of-the-art sequential implementation of recompression in
  uncompressed $\bigO(\Textlen)$ time. When enabling parallelism
  in~\cite{recompression-github}, our variant is still at least
  $18\times$ faster.

\item \emph{Massive RAM reduction}: On the same 4\,GiB file as above,
  our implementation uses $17\times$ less RAM. When compared to the
  parallel implementation from~\cite{recompression-github}, we use
  $14\times$ less RAM.

\end{itemize}

This improved scalability translates not only to time and RAM usage
gains but also to the ability to process larger datasets. For
example, when attempting to run our experiments on an 8\,GiB test
file, the state-of-the-art implementation always runs out of RAM on
our 94\,GiB system, resulting in significantly worse performance.

Our implementation confirms the enormous potential of computation in
compressed time for the construction of compressed text
indexes~\cite{resolution,collapsing}. All our implementations are
available at
\url{https://github.com/AnkithReddy02/fast-recompression}.

\section{Preliminaries}\label{sec:prelim}

\paragraph{Strings}

For any string $S$, we write $S[i \dd j]$, where $1 \leq i, j \leq
|S|$, to denote the substring of $S$ starting at position $i$ and
ending at $j$. If $i > j$, we define $S[i \dd j]$ to be the empty
string $\emptystring$. By $[i \dd j)$ we denote $[i \dd
j-1]$. Throughout the paper, we consider a string (text) $\Text[1 \dd
\Textlen]$ of length $\Textlen \geq 1$ over an integer alphabet
$\Sigma = \IntegerAlphabet$. The function $\LCE{\Text}{i}{i'}$ denotes
the length of the longest common prefix of $\Text[i \dd
\Textlen]$ and $\Text[i' \dd \Textlen]$.

\paragraph{LZ77 Compression}

An \emph{LZ77-like factorization} of a string $\Text$ is a partition
$\Text = F_1 \cdots F_f$ into non-empty \emph{phrases} such that each
phrase $F_j$ with $|F_j| > 1$ appears earlier in $\Text$, i.e.,
letting $i = 1 + |F_1 \cdots F_{j-1}|$ and $\ell = |F_j|$, there
exists $p \in [1 \dd i)$ such that $\LCE{\Text}{p}{i} \geq \ell$. The
phrase $F_j = \Text[i \dd i + \ell)$ is represented by the pair $(p,
\ell)$. If multiple values of $p$ are valid, one is chosen
arbitrarily. The segment $\Text[p \dd p + \ell)$ is called the
\emph{source} of $F_j$. If $|F_j| = 1$, then we do not require that
$F_j = \Text[i]$ occurs earlier. Such phrases are represented as
$(\Text[i], 0)$.

The \emph{LZ77 factorization}~\cite{LZ77} (or \emph{LZ77 parsing}) of a string
$\Text$ is an LZ77-like factorization obtained by greedily scanning
$\Text$ from left to right and selecting the longest possible
phrases. More precisely, the $j$th phrase $F_j$ is the longest
substring starting at position $i = 1 + |F_1 \cdots F_{j-1}|$ that has
a prior occurrence in $\Text$. If no such substring exists, then $F_j
= \Text[i]$. The number of phrases in the LZ77 parsing is denoted by
$\LZSize{\Text}$. For example, the string
$\texttt{bbabaababababaababa}$ has the LZ77 parsing $\texttt{b} \cdot
\texttt{b} \cdot \texttt{a} \cdot \texttt{ba} \cdot \texttt{aba} \cdot
\texttt{bababa} \cdot \texttt{ababa}$, with $\LZSize{\Text} = 7$
phrases, represented by the sequence $(\texttt{b},0), (1,1),
(\texttt{a},0), (2,2), (3,3), (7,6), (10,5)$.

The \emph{non-overlapping LZ77 factorization} is defined similarly,
except that the previous occurrence of each phrase must not overlap
the phrase itself. The number of phrases in this variant is denoted
$\LZNonOvSize{\Text}$.

\paragraph{Grammar Compression}

A context-free grammar is a tuple $G = (N, \Sigma, R, S)$, where $N$
is a finite set of \emph{nonterminals}, $\Sigma$ is a finite set of
\emph{terminals}, and $R \subseteq N \times (N \cup \Sigma)^*$ is a
set of \emph{rules}. We assume $N \cap \Sigma = \emptyset$ and $S \in
N$. The symbol $S$ is called the \emph{start symbol}. If $(A,
\gamma) \in R$, we write $A \rightarrow \gamma$. The \emph{language}
of $G$ is the set $L(G) \subseteq \Sigma^*$ obtained by starting from
$S$ and repeatedly replacing nonterminals according to $R$.

A grammar $G = (N, \Sigma, R, S)$ is called a \emph{straight-line
grammar} (SLG) if each $A \in N$ appears in exactly one production
on the left-hand side, and the nonterminals can be ordered as $A_1,
\ldots, A_{|N|}$ such that, whenever $A_i \rightarrow
\gamma$, it holds that $\gamma \in (\Sigma \cup \{A_{i+1}, \ldots,
A_{|N|}\})^*$. This ensures that the grammar rule graph is
acyclic. The unique $\gamma$ such that $A \rightarrow \gamma$ is
called the \emph{definition} of $A$, and denoted $\Rhs{G}{A}$. For any $u
\in (N \cup \Sigma)^*$, there is exactly one $w \in \Sigma^*$
derivable from $u$. We call such $w$ the \emph{expansion} of $u$,
and denote it $\Exp{G}{u}$.

The principle of \emph{grammar compression} involves constructing, for
a given text $\Text$, a small SLG $G$ such that $L(G) =
\{\Text\}$. The \emph{size} of the grammar is the total length of all
definitions: $|G| := \sum_{A \in N}|\Rhs{G}{A}|$.

A \emph{straight-line program (SLP)} is an SLG in which each
production is either $A \rightarrow XY$ with $X,Y \in N$, or $A
\rightarrow c$ with $c \in \Sigma$. Every SLG $G$ can be converted in
$\bigO(|G|)$ time into an SLP $G'$ representing the same~string.

A \emph{run-length SLP (RLSLP)} is a generalized SLP that also allows
rules of the form $A \rightarrow X^k$, where $X \in N$ and $k \in
\Zp$. For each such nonterminal, we define $|\Rhs{G}{A}| = 2$. The
size of an RLSLP is defined analogously to that of SLPs.

\begin{theorem}[{\cite{Rytter03,charikar}}]\label{th:nonov-lz77-to-slp}
  Given the non-overlapping LZ77 factorization of a string $\Text$ of
  length $\Textlen$, we can in $\bigO(\LZNonOvSize{\Text} \log
  \Textlen)$ time construct an SLP $G$ such that $\Lang{G} =
  \{\Text\}$.
\end{theorem}

\begin{theorem}[{\cite[Lemma\,8]{Gawrychowski11}, \cite[Theorem\,6.1]{resolution}}]\label{th:lz77-to-slp}
  Given the LZ77 factorization of a string $\Text$ of length
  $\Textlen$, we can in $\bigO(\LZSize{\Text} \log \Textlen)$ time
  construct an SLP $G$ such that $\Lang{G} = \{\Text\}$.
\end{theorem}

\begin{theorem}[{\cite[Proposition~3.2]{collapsing}}]\label{th:lz77-like-to-lz77}
  Given a string $\Text$ of length $\Textlen$, represented using an
  LZ77-like parsing consisting of $f$ phrases, the LZ77 parsing of
  $\Text$ can be constructed in $\bigO(f \log^4 \Textlen)$ time.
\end{theorem}

\section{Overview of Recompression}\label{sec:recompression}

Recompression is a lossless compression algorithm introduced by
Jeż~\cite{Jez15,Jez16} that builds a run-length straight-line program
(RLSLP) by repeatedly applying local rewriting rules to a string. In
each recompression round, the input string is reduced using two key
operations:

\begin{itemize}

\item \textbf{Block Compression (BComp).} This step replaces every
  maximal block of at least two identical symbols with a fresh
  nonterminal, ensuring that adjacent characters in the resulting
  string are distinct. For example, applying BComp to $\Text =
  \texttt{abaaabccaaa}$ produces two new rules, $X \rightarrow
  \texttt{a}^3$ and $Y \rightarrow \texttt{c}^2$, and replaces $\Text$
  with $\texttt{ab}X\texttt{b}YX$.

\item \textbf{Pair Compression (PComp).} This step replaces frequent
  pairs of adjacent symbols with new nonterminals. The algorithm
  partitions the current alphabet into two disjoint sets: a left set
  $\Sigma_L$ and a right set $\Sigma_R$. Only pairs $ab$ with $a \in
  \Sigma_L$ and $b \in \Sigma_R$ are replaced. For example, applying
  PComp to $\Text = \texttt{abacaabad}$ with $\Sigma_{L} =
  \{\texttt{a}, \texttt{c}\}$ and $\Sigma_{R} = \{\texttt{b},
  \texttt{d}\}$ creates two new rules, $X \rightarrow \texttt{ab}$ and
  $Y \rightarrow \texttt{ad}$, and $\Text$ is replaced with
  $X\texttt{aca}XY$. The computation of the partition $\Sigma =
  \Sigma_L \cup \Sigma_R$ is performed to maximize the number of
  replaced pairs (i.e., pairs $ab$ in $\Text$ with $a \in
  \Sigma_L$ and $b \in \Sigma_R$). One simple strategy is to assign
  every $c \in \Sigma$ to either $\Sigma_L$ or $\Sigma_R$
  randomly. The expected number of reduced pairs is then
  $\tfrac{1}{4}(|\Text|-1)$. This strategy can be derandomized,
  resulting in a linear-time partitioning algorithm that
  \emph{guarantees} at least one quarter of all adjacent symbol pairs
  are replaced. This ensures that the string shrinks by a constant
  factor~\cite{Jez15,Jez16}.

\end{itemize}

The BComp and PComp steps are applied alternately in rounds, resulting
in a sequence of strings $\Text_0, \Text_1, \Text_2, \dots$ such that
$\Text_0 = \Text$, and for every $i > 0$, $\Text_i$ is obtained from
$\Text_{i-1}$ by applying BComp (if $i$ is odd) or PComp (if $i$ is
even). The process stops when the string reduces to a single symbol,
i.e., when $|\Text_k| = 1$ for some $k$. For a string of length
$\Textlen$, each BComp and PComp round runs in $\bigO(\Textlen)$ time
and space. Since the string length is reduced by a constant factor
every two rounds, the overall algorithm runs in $\bigO(\Textlen)$ time
and uses $\bigO(\Textlen)$ space~\cite{Jez15,Jez16}.

A complete example of the recompression process, including
round-by-round transformations, can be found in~\cite{Jez15,Jez16}.
We refer the reader to that work for additional technical details and
formal analysis.

\paragraph{Recompression in Compressed Time}

In the above scenario, we assumed that the recompression algorithm was
applied to an uncompressed text $\Text$ of length $\Textlen$. However, the main
goal of our paper is to implement the recompression algorithm directly
on the SLP-compressed input. In~\cite{Jez15,Jez16}, Jeż describes how
to achieve this. More precisely, the paper shows that, given an SLP
$G$ of size $|G| = g$, one can perform recompression of the text $\Text$
(obtained by decompressing $G$) in only $\bigO(g \log \Textlen)$ time.

\section{Implementation Details}\label{sec:implementation}

Our implementation is modeled after the description of a variant of
recompression by Kempa and Kociumaka in~\cite{collapsing},
which achieves the optimal space
bound of\footnote{The
  $\bigO\!\left(\SubstringComplexity{\Text}\log \frac{\Textlen \log
  \AlphabetSize}{\SubstringComplexity{\Text}\log \Textlen}\right)$
  bound is known to be the smallest possible space to represent a
  string $\Text \in \IntegerAlphabet^{\Textlen}$ for every
  combination of $\Textlen$, $\AlphabetSize$, and
  $\SubstringComplexity{\Text}$~\cite{delta}.}
\[
  \bigO\!\left(\SubstringComplexity{\Text}\log \frac{\Textlen \log
    \AlphabetSize}{\SubstringComplexity{\Text}\log \Textlen}\right)
\]
(where the substring complexity $\SubstringComplexity{\Text}\leq
\LZSize{\Text}$ is a fundamental measure of
compressibility~\cite{delta}), although we did not
implement the more intricate optimizations needed to achieve this space,
and the result of our computation is still the original recompression
grammar, as defined in~\cite{Jez15,Jez16}. More specifically, in our
implementation we follow Section~5.2 of~\cite{collapsing}—notably
Definitions~5.2 and~5.4, Lemmas~5.3 and~5.5, and Construction~5.6.

In this section, we describe the practical techniques we employed
in our implementation to achieve small runtime and low peak RAM
usage.

\subsection{Representing Intermediate Strings}\label{sec:implementation-representation}

In the construction from~\cite[Section~5.2]{collapsing}, the strings
$\Text_0, \Text_1, \Text_2, \dots$ during recompression are
represented using their LZ77 factorization. When computing $\Text_{i}$
from $\Text_{i-1}$, the algorithm first converts the LZ77
factorization of $\Text_{i-1}$ into an SLP $G_{i-1}$ in
$\bigO(\LZSize{\Text_{i-1}} \log \Textlen)$ time
(Theorem~\ref{th:lz77-to-slp}). In $\bigO(|G_{i-1}|)$ time it then performs
either BComp or PComp on the resulting SLP. The result of the
computation is an SLG $G'_i$ representing $\Text_{i}$. This SLG is
then converted in $\bigO(|G'_i|)$ time into an LZ77-like factorization
of $\Text_{i}$ and finally into the LZ77 factorization of $\Text_{i}$
using Theorem~\ref{th:lz77-like-to-lz77}.

One of the changes in our implementation, compared to~\cite{collapsing},
is that each of the intermediate strings
$\Text_0, \Text_1, \Text_2, \dots$ during recompression is represented as an
SLG. This affects the implementation as follows:
\begin{itemize}
\item Adopting the SLG representation slightly
  changes the implementation of BComp and PComp, since the right-hand side
  of a nonterminal can now be longer than two symbols. However, this way we
  avoid the conversions to and from the LZ77 factorization.
\item Storing an SLG in small space is more challenging than storing
  an SLP, since definitions of nonterminals are variable-length
  strings. In our implementation we store the definitions of all
  nonterminals in a single array. Each nonterminal then stores a single
  integer indicating the starting position of its definition. This reduces
  memory overhead while supporting fast access to every
  definition.
\end{itemize}

\subsection{Block Compression (BComp)}\label{sec:implementation-bcomp}

The general approach to performing the BComp procedure over a text
$\Text_{i-1}$ represented using an SLG $G$ is to perform a bottom-up
construction of another SLG $G'$ representing the run-length
compressed string; see Lemma~5.3 in~\cite{collapsing}.

\paragraph{Computing Left and Right Runs}

The construction of grammar $G'$ requires computing,
for each nonterminal $A$, the maximal prefix and suffix
blocks that consist of the same symbol. These are denoted
$\mathsf{LR}(A) = (X, \ell)$ and $\mathsf{RR}(A) = (Y, r)$,
representing runs of length $\ell$ and $r$, respectively. To optimize
storage, we use two arrays:
\[
  \texttt{LR\_vec}[A] = (X, \ell), \qquad \texttt{RR\_vec}[A] = (Y, r)
\]
where $\ell, r < 255$ are stored as 8-bit values packed alongside
the symbol ID. In rare cases where $\ell$ or $r$ are at least
255, we store them in auxiliary arrays \texttt{large\_LR\_vec}
and \texttt{large\_RR\_vec}. In this case, we set the corresponding
8-bit value to 255 and use binary search to retrieve the actual count.

\paragraph{Supporting Grammar Updates}

During the construction of grammar $G'$, we need to add new nonterminals
and their definitions. An important property of the construction of $G'$
is that it never deletes or changes any already added rule of $G'$.
This lets us implement dynamic updates
in $G'$ in a relatively straightforward way, i.e., the array storing the
definitions of all nonterminals is implemented as a standard dynamic array
that supports insertion at the end.

Our implementation uses a custom dynamic array (which we
call a \texttt{space\_efficient\_vector}—a drop-in
replacement for standard \texttt{C++} solutions like
\texttt{std::vector}) that stores elements in a bounded number of
dynamically allocated blocks. Each block's size doubles only when all
existing blocks are full, so growth requires reallocating and copying
at most two adjacent blocks at a time rather than the entire array.
This chunked strategy keeps the logical interface of a random-access
array while reducing peak memory during expansions, since no
full duplicate of the data is ever created.

\subsection{Pair Compression (PComp)}\label{sec:implementation-pcomp}

The implementation of the PComp procedure, which, given the SLG $G$
representing the string $\Text_{i-1}$, computes the SLG $G'$
representing $\Text_i$ (i.e., the result of applying PComp to
$\Text_{i-1}$), is slightly more complex than BComp, since it
needs to first compute a good partitioning of the alphabet $\Sigma$ into
$\Sigma_{L}$ and $\Sigma_{R}$, and then perform the transformation of
$\Text_{i-1}$ that replaces every pair of symbols $ab$, with $a \in
\Sigma_{L}$ and $b \in \Sigma_{R}$, by a fresh nonterminal.

The second step is similar to BComp and reduces to processing all
nonterminals of $G$ bottom up; see Lemma~5.5 of~\cite{collapsing}.

The more complex step of PComp is the computation of the alphabet
partitioning. In our implementation, we employ two strategies, which
alternate in subsequent executions of PComp.

\begin{itemize}
\item The randomized partition simply assigns terminals to the left or
  right sets uniformly at random.
\item The deterministic partition uses a MaxCut-inspired greedy
  strategy that guarantees the string
  shrinks by at least 25\% after applying PComp~\cite{Jez15}.
\end{itemize}

Implementing the random partition is straightforward. One of the key steps to
implement the deterministic partition is to compute the frequency of
every pair $ab$ in the string $\Text_{i-1}$ represented using the SLG
$G$. It is known (see, e.g.,~\cite[Lemma~3]{charikar}) that the number
of such distinct pairs is bounded by $2|G|$, and the computation can
be done in $\bigO(|G|)$ time (see,
e.g.,~\cite[Lemma~15]{tomohiro-lce}). The computation of the partition
$\Sigma_{L}$ and $\Sigma_{R}$ is then a straightforward adaptation of
the MaxCut approximation algorithm (see,
e.g.,~\cite[Lemma~5.19]{collapsing}
or~\cite[Algorithm~1]{tomohiro-lce}).

\section{Experiments}\label{sec:experiments}

\subsection{Setup}

All experiments were conducted on a machine equipped with two Intel
Xeon X5690 CPUs, each operating at 3.47\,GHz, totaling 12 physical cores.
The system had 94\,GiB of DDR3
RAM and 931\,GiB of local disk space. Disk
throughput was measured at approximately 102\,MiB/s. The
operating system was Ubuntu 16.04.7 running Linux kernel
4.15.0. All programs were compiled with \texttt{g++}
version 5.4.0 using the flags \texttt{-funroll-loops -O3 -DNDEBUG
  -march=native -std=c++17 -pthread}.
All runtimes are reported as wall-clock times.

\subsection{Datasets}\label{sec:datasets}

We evaluate our algorithms on highly repetitive datasets derived from the
\emph{Saccharomyces cerevisiae} genome. Specifically, we use the
\texttt{chr16.fsa} file (941.4\,KiB) from the Saccharomyces Genome Database (SGD)
database\footnote{\url{http://sgd-archive.yeastgenome.org/sequence/S288C_reference/chromosomes/fasta/}}
as the base input. Larger instances are generated by repeated
duplication of the sequence followed by uniform random mutations at
rates $10^{-4}$ and $10^{-5}$ per symbol. Basic statistics of our datasets
are listed in Table~\ref{tab:datasets}.

Each 4\,GiB dataset is a strict prefix of its corresponding 8\,GiB
version, allowing fair comparison across scales while preserving
structure. We refer to each dataset using the shorthand
\textbf{Y\textit{S}M\textit{E}}, where \textit{S} is the input size in
GiB and \textit{E} denotes the exponent in the mutation rate
$10^{-E}$. For example, \textbf{Y4M5} refers to a 4\,GiB input with a
mutation rate of $10^{-5}$. Unless otherwise specified, experiments
default to \textbf{Y4M5}.

\begin{table}[t]
  \small
  \centering
  \begin{tabular}{@{}cccr@{}}
  \toprule
  \textbf{Name} & \boldmath$n/2^{30}$ & \boldmath$|\Sigma|$ & \boldmath$n/z$ \\
  \midrule
  Y4M4 & 4 & 4 & 4{,}625 \\
  Y4M5 & 4 & 4 & 23{,}087 \\
  Y8M4 & 8 & 4 & 5{,}043 \\
  Y8M5 & 8 & 4 & 30{,}990 \\
  \bottomrule
  \end{tabular}
  \caption{Statistics of datasets used in our experiments, with $n$ denoting
    the string length, $|\Sigma|$ the alphabet size, and $n/z$ the average phrase
    length in the (exact) LZ77 factorization.\protect\footnotemark\ The strings
    used in our experiments do not use symbol packing, i.e., each symbol is encoded
    using a single byte.}\label{tab:datasets}
\end{table}

\footnotetext{We computed the value $n / z$ for our datasets
  using the \texttt{KKP2n} algorithm~\cite{lazylz}. Its source code is
  available at \url{https://github.com/dominikkempa/lazy-lz77}.}

\subsection{Implementations}\label{sec:implementations}

In our experiments, we evaluate two categories of algorithms for computing
a recompression RLSLP.

\begin{itemize}
\item \textbf{Standard (uncompressed) recompression:} We use the recompression
  implementation by Osthues~\cite{recompression-github}, available at
  \url{https://github.com/christopherosthues/recompression}. This
  implementation performs recompression directly on the input string
  without prior LZ parsing. We evaluate the following variants:

  \begin{itemize}
  \item \texttt{fast\_seq}: A sequential implementation that
    alternates between block compression (on repeated symbols) and
    pair compression (on disjoint symbol pairs).

  \item \texttt{parallel\_rnd}: A parallel variant that randomly
    partitions the alphabet during pair compression, improving
    parallel scalability at the cost of nondeterminism.

  \item \texttt{parallel\_gr}: A parallel method that applies a greedy
    MaxCut-based partitioning strategy to deterministically improve
    compression quality compared to the randomized variant.
  \end{itemize}

\item \textbf{Recompression in compressed time (this work):}
  Our pipeline performs compression in five stages:

  \begin{enumerate}
  \item \textbf{Text $\rightarrow$ Approximate LZ77:} Approximate
    LZ77 parsing is computed using the Bentley--McIlroy
    algorithm~\cite{bentley1999dcc}.\footnote{We use the term \emph{approximate LZ77 parsing}
    as an alternative to \emph{LZ77-like parsing}. The approximate LZ77 parsing still encodes
    the text $\Text$ exactly, i.e., we can restore $T$ without any errors.} Our implementation
    of the Bentley--McIlroy algorithm is available at
    \url{https://github.com/AnkithReddy02/fast-recompression}.

  \item \textbf{Approximate LZ77 $\rightarrow$ SLG:} The approximate LZ77
    factorization is converted into a straight-line grammar (SLG)
    using the lazy AVL grammar construction described
    in~\cite{KempaL21}. This implementation is available at
    \url{https://github.com/dominikkempa/lz77-to-slp}.

  \item \textbf{SLG $\rightarrow$ SLP:} The SLG is expanded into a
    full straight-line program (SLP) generating the original string.
    Our tool for this task is available at
    \url{https://github.com/AnkithReddy02/fast-recompression}.

  \item \textbf{SLP $\rightarrow$ Pruned SLP:} Unused rules are
    eliminated via pruning without affecting the represented string.
    Our implementation of this pruning step is also available at
    \url{https://github.com/AnkithReddy02/fast-recompression}.

  \item \textbf{Pruned SLP $\rightarrow$ RLSLP:} A recompression
    RLSLP is constructed in compressed time from the SLP-compressed
    input text, as described in Section~\ref{sec:implementation}.
    This implementation constitutes the main contribution of our paper
    and is available at
    \url{https://github.com/AnkithReddy02/fast-recompression}.
  \end{enumerate}

  During the \texttt{Pruned SLP} $\rightarrow$ \texttt{RLSLP} stage,
  we evaluate three partitioning strategies for pairwise compression:
  deterministic, randomized, and mixed. The deterministic strategy
  applies a greedy MaxCut-based heuristic to partition the
  alphabet; the randomized strategy selects partitions uniformly at
  random; and the mixed strategy alternates between deterministic and
  randomized partitioning at each pairwise compression round. Unless
  otherwise stated, we use the mixed strategy. A
  comparison of the three strategies is presented in
  Section~\ref{sec:experiment5}.

  The correctness of the recompression RLSLPs produced by our
  compressed pipeline was verified by decompressing them and comparing
  the results with the original texts. To verify the local consistency
  property, we implemented longest-common-extension (LCE) queries on
  the recompression RLSLPs (as described
  in~\cite[Theorem~5.25]{collapsing}) and compared their outputs with
  those from an independent LCE implementation.
\end{itemize}

The peak RAM usage in all implementations was measured
using manual tracking of memory allocations. In the case of the
uncompressed versions, we used the provided tracking mechanism. All
tools in our compressed pipeline use our custom memory tracking.

\subsection{Experiment 1: Measuring Overall Time to Construct the Recompression RLSLP}\label{sec:experiment1}

In the first experiment, we evaluate the end-to-end performance of our
pipeline for constructing a recompression RLSLP in compressed time.
We compare it against three state-of-the-art
uncompressed baselines (\texttt{fast\_seq}, \texttt{parallel\_gr},
and \texttt{parallel\_rnd}).

We run all methods on the four datasets described in
Section~\ref{sec:datasets} and report runtime, peak memory usage, and
the number of productions in the resulting recompression RLSLP in
Table~\ref{tab:recompression-results}. Our compressed method is
evaluated with block sizes 50 and 500 (a parameter in the
Bentley--McIlroy algorithm controlling the trade-off between
time/space and output size). We tested a wider range of values and
found these two to represent a good balance.

\begin{table*}[t]
    \centering
    \small
    \begin{tabular}{@{}lcrrr@{}}
    \toprule
    \textbf{Dataset} & \textbf{Algorithm} & \textbf{Runtime (s)} & \textbf{Peak RAM (GiB)} & \textbf{RLSLP Productions} \\
    \midrule
    Y4M4 & Compressed (50) & 109.61 & 4.06 & 5,307,093 \\
    Y4M4 & Compressed (500) & 171.48 & 4.25 & 5,332,128 \\
    Y4M4 & fast\_seq & 1394.00 & 70.56 & 4,890,650 \\
    Y4M4 & parallel\_gr & 529.00 & 58.38 & 6,036,842 \\
    Y4M4 & parallel\_rnd & 484.00 & 58.38 & 6,084,384 \\
    \midrule
    Y4M5 & Compressed (50) & 41.99 & 4.02 & 936,193 \\
    Y4M5 & Compressed (500) & 24.45 & 4.03 & 934,699 \\
    Y4M5 & fast\_seq & 1134.00 & 70.56 & 864,360 \\
    Y4M5 & parallel\_gr & 520.00 & 58.38 & 1,046,708 \\
    Y4M5 & parallel\_rnd & 457.00 & 58.38 & 1,055,740 \\
    \midrule
    Y8M4 & Compressed (50) & 241.83 & 8.09 & 8,998,801 \\
    Y8M4 & Compressed (500) & 365.52 & 8.44 & 9,038,760 \\
    Y8M4 & fast\_seq & 4674.00 & 137.83 & 8,953,448 \\
    Y8M4 & parallel\_gr & 2594.00 & 114.06 & 10,903,809 \\
    Y8M4 & parallel\_rnd & 2366.00 & 114.06 & 10,967,268 \\
    \midrule
    Y8M5 & Compressed (50) & 101.81 & 8.02 & 1,521,123 \\
    Y8M5 & Compressed (500) & 48.16 & 8.04 & 1,508,313 \\
    Y8M5 & fast\_seq & 3750.00 & 137.83 & 1,526,272 \\
    Y8M5 & parallel\_gr & 2560.00 & 114.06 & 1,853,228 \\
    Y8M5 & parallel\_rnd & 2633.00 & 114.06 & 1,866,394 \\
    \bottomrule
    \end{tabular}
    \caption{Comparison of uncompressed and compressed approaches
      for constructing the recompression RLSLP.}
    \label{tab:recompression-results}
\end{table*}

\paragraph{Key Observations}
\begin{itemize}

\item \textbf{Drastic RAM Reduction}\, The peak RAM usage of the
  compressed pipeline scales linearly with the input size: from
  4.02--4.25\,GiB for 4\,GiB inputs to 8.02--8.44\,GiB for 8\,GiB
  inputs. In contrast, the uncompressed methods require
  58.38--70.56\,GiB on 4\,GiB inputs and 114.06--137.83\,GiB on 8\,GiB
  inputs, corresponding to a memory overhead of 13--17$\times$.

  The memory usage of uncompressed variants not only affects space
  requirements but also performance. On the 8\,GiB datasets, all
  uncompressed methods exceed the 94\,GiB memory limit of the system
  and trigger swap usage, whereas the compressed pipeline runs fully
  in memory.

\item \textbf{Massive Runtime Gains}\, Our compressed pipeline
  delivers its most striking gains on large, highly repetitive
  datasets, which are the primary focus of this work. Against the
  sequential uncompressed recompression (\texttt{fast\_seq}),
  the advantage is overwhelming: on Y4M5 our \texttt{Compressed (500)}
  configuration finishes in 24\,s versus 1,134\,s, a 47$\times$
  speedup; on Y8M5 the gap widens to 48\,s versus 3,750\,s, or
  78$\times$ faster.\footnote{This increased gap is likely at least
  partially due to the uncompressed implementation resorting to swap
  (disk) memory, and hence we accept the 47$\times$ speedup as our maximal
  reliable speedup.}
  Even when compared to \emph{parallel}
  uncompressed implementations (which benefit from multiple cores),
  our single-threaded approach remains far ahead: parallel variants
  need at least 457\,s on 4\,GiB inputs and at least 2,560\,s on 8\,GiB
  inputs, yielding speedups of 19$\times$ and 53$\times$.

  Even for the less repetitive datasets (Y4M4 and Y8M4), the benefits
  of compressed computation remain substantial; e.g.,
  \texttt{Compressed (500)} requires 171\,s on Y4M4 and 366\,s on
  Y8M4, cutting sequential runtime by roughly 8$\times$--13$\times$
  and staying faster than every parallel uncompressed variant despite
  their multicore advantage.

\item \textbf{Stable Output Size}\, Our compressed pipeline produces
  RLSLPs that are either smaller or only slightly larger (but never
  by more than 10\%) than the sequential uncompressed implementation
  of recompression. Furthermore, our RLSLPs are always smaller than
  those produced by the parallel variants of uncompressed
  recompression.

\end{itemize}

\begin{figure*}[t]
  \centering
  \begin{subfigure}[t]{0.48\textwidth}
    \includegraphics[width=\linewidth]{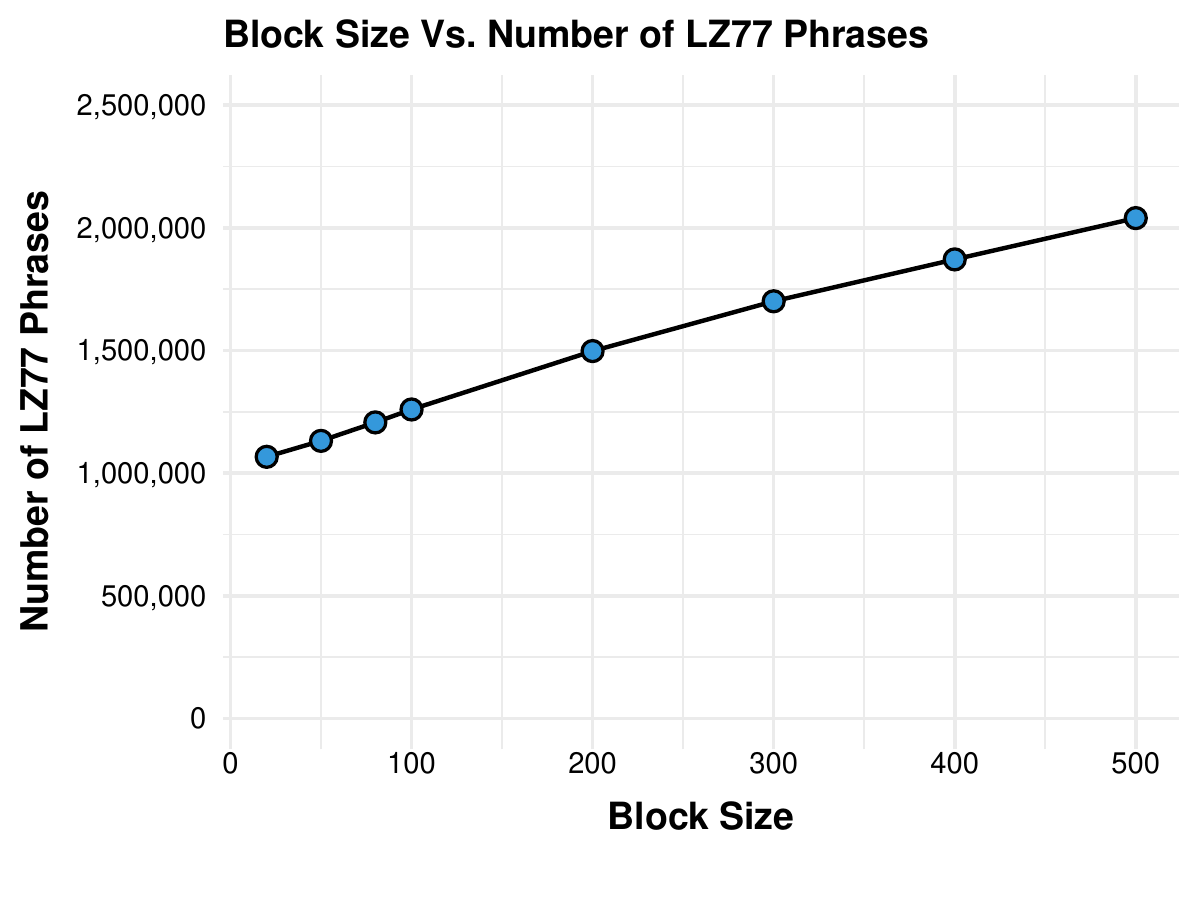}
    \caption{Number of LZ77 phrases}
    \label{fig:lz77phrases}
  \end{subfigure}
  \hfill
  \begin{subfigure}[t]{0.48\textwidth}
    \includegraphics[width=\linewidth]{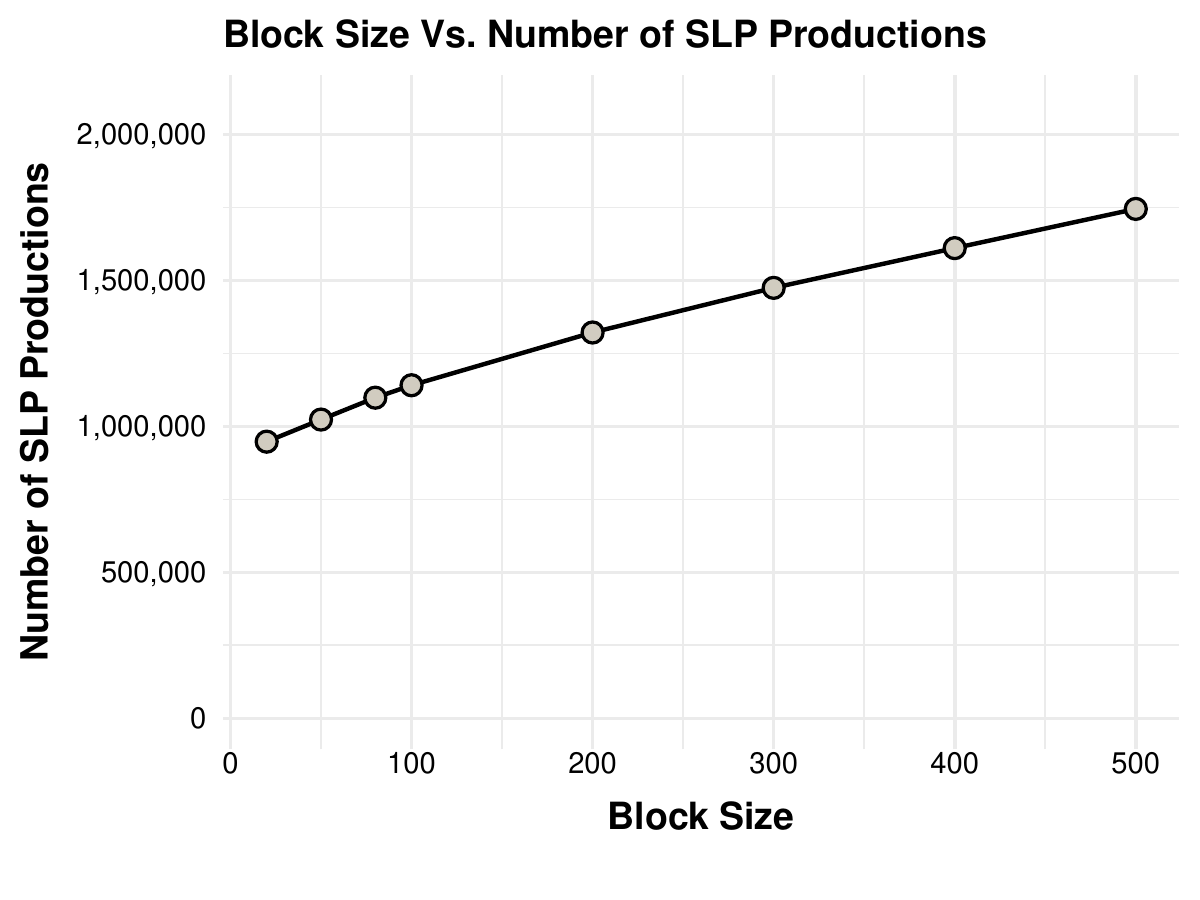}
    \caption{Number of SLP productions}
    \label{fig:slpsize}
  \end{subfigure}
    
  \vspace{1em}
    
  \begin{subfigure}[t]{0.48\textwidth}
    \includegraphics[width=\linewidth]{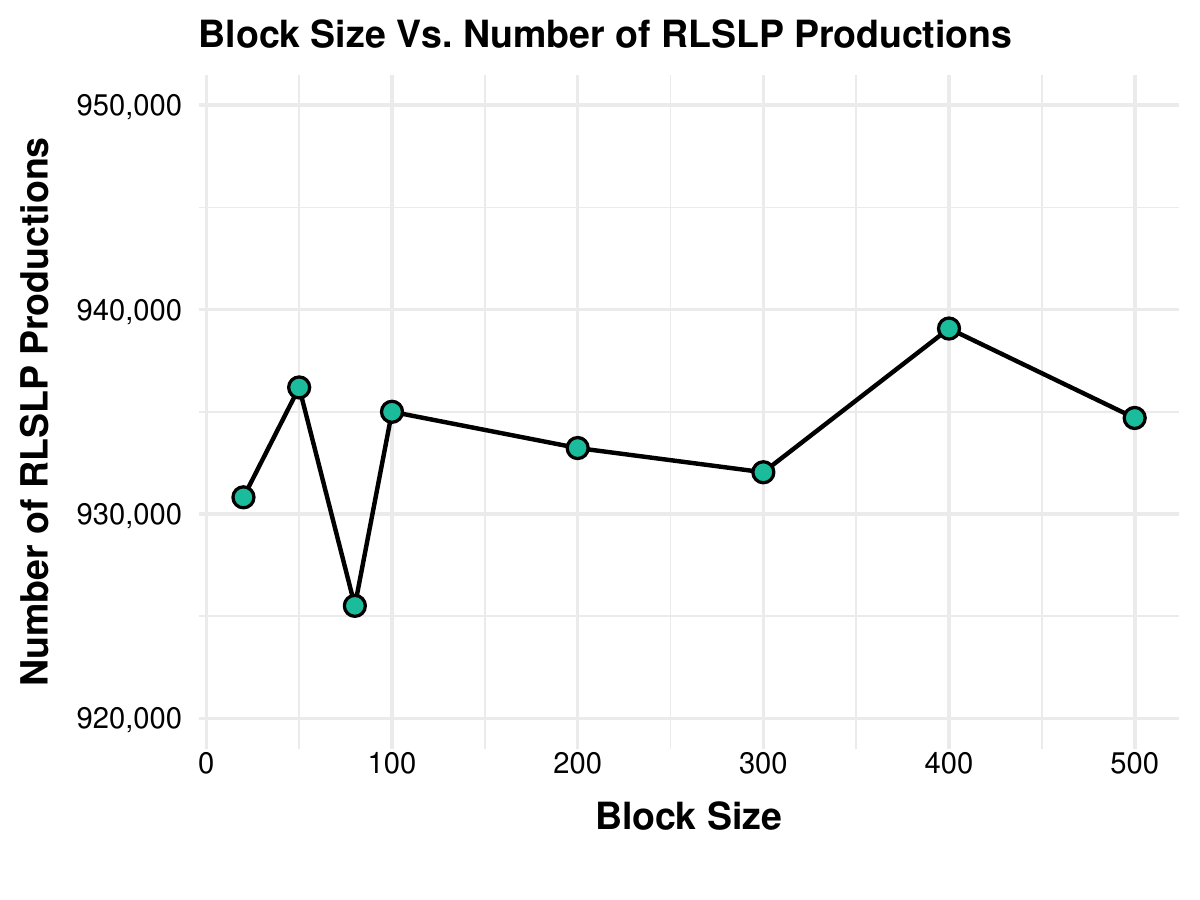}
    \caption{Number of RLSLP productions}
    \label{fig:rlslpsize}
  \end{subfigure}
  \hfill
  \begin{subfigure}[t]{0.48\textwidth}
    \includegraphics[width=\linewidth]{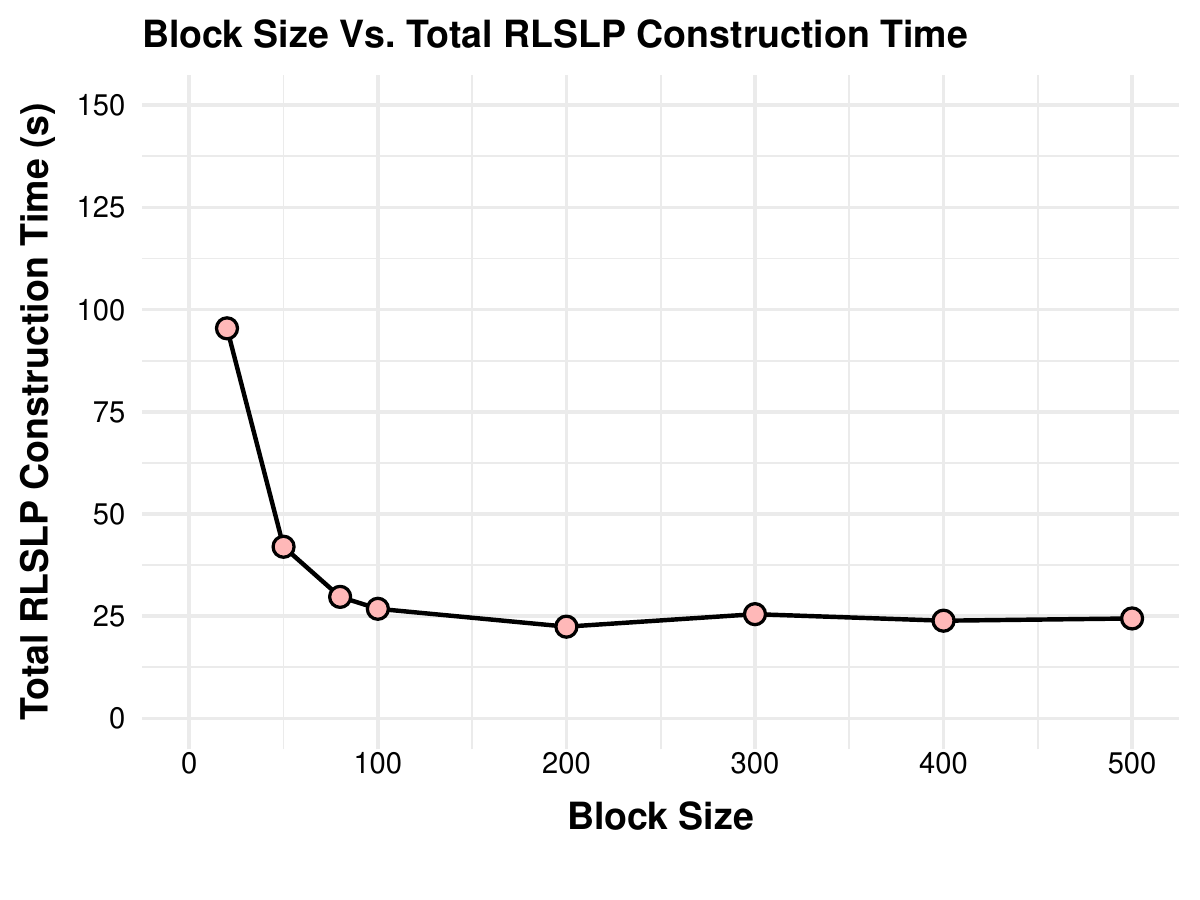}
    \caption{Pipeline runtime}
    \label{fig:runtime}
  \end{subfigure}
    
  \caption{Impact of block size on compression pipeline components
    (dataset: Y4M5).}
  \label{fig:exp1}
\end{figure*}

\subsection{Experiment 2: Effect of Block Size in LZ Approximation}\label{sec:experiment2}

This experiment examines the effect of block size on intermediate
metrics within the compressed pipeline. We report the number of
phrases in the resulting LZ77-like parsing,
the number of productions in the pruned SLP,
the number of productions in the final recompression RLSLP, and the overall
runtime. All results are computed for the Y4M5 dataset
(we also ran the experiment on Y8M5 and found the trends
to be very similar to Y4M5).
Results are shown in Figures~\ref{fig:lz77phrases}--\ref{fig:runtime}.

\paragraph{Number of Phrases}

Figure~\ref{fig:lz77phrases} shows that the number of phrases
increases with block size. At block size~20, the parser emits roughly
1.1 million phrases, increasing to over 2 million at block size~500.

\paragraph{Number of SLP Productions}

As shown in Figure~\ref{fig:slpsize}, the number of productions in the
pruned SLP correlates closely with the phrase count. The smallest SLP occurs
at block size~20, while the largest occurs at block size~500.

\paragraph{Number of RLSLP Productions}

Figure~\ref{fig:rlslpsize} indicates that the number of productions in the
resulting recompression RLSLP remains stable across all block sizes,
fluctuating within the narrow range of 925,000--940,000 productions.
This stability suggests that downstream compression is effective in
absorbing variation introduced by the LZ77 approximation.

\paragraph{Pipeline Runtime}

As shown in Figure~\ref{fig:runtime}, runtime decreases substantially
as block size increases. At block size~20, the full pipeline takes
nearly 100\,s. Beyond block size~100, the runtime flattens, with block
sizes 100 and 500 both completing in approximately 25\,s. The largest gains
come from reducing parsing overhead at small block sizes.

These results highlight a trade-off between compression and speed:
smaller blocks allow longer matches, producing fewer phrases and
slightly smaller intermediate grammars, but they sharply increase the
cost of parsing. When the pipeline starts \emph{from raw text}, that
extra effort brings little benefit: the downstream recompression stage
quickly reduces the grammar to nearly the same size regardless of
block size. The size of the initial LZ77-like parsing is not,
however, irrelevant. As shown in Experiment~4, if a high-quality
LZ77-like parse is \emph{already available} --- for example, computed
on another machine --- the peak RAM of the entire pipeline
scales with the size of that parse, so a smaller, more refined
factorization can reduce memory use.

\begin{figure*}[t]
  \begin{minipage}{0.5\textwidth}
    \includegraphics[width=\columnwidth]{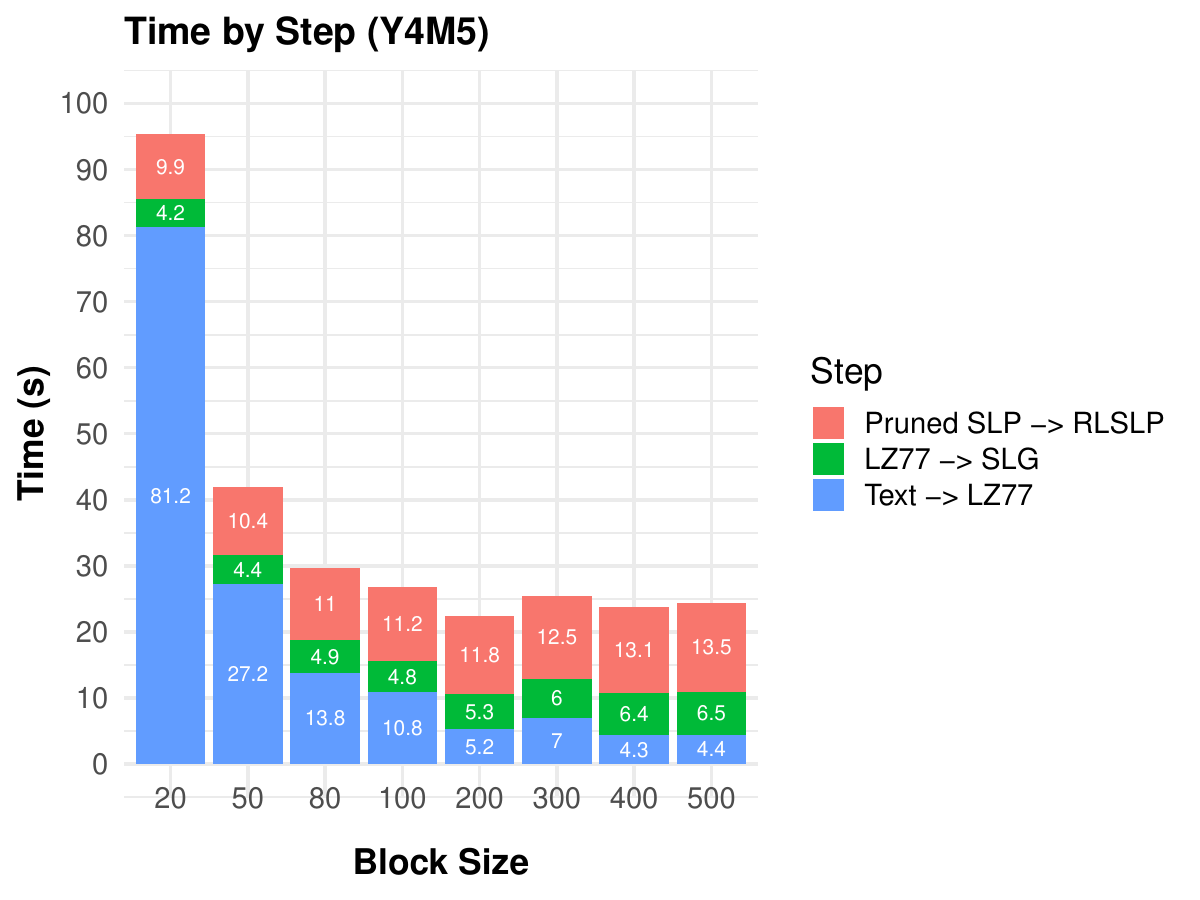}
  \end{minipage}
  \hfill
  \begin{minipage}{0.5\textwidth}
    \includegraphics[width=\columnwidth]{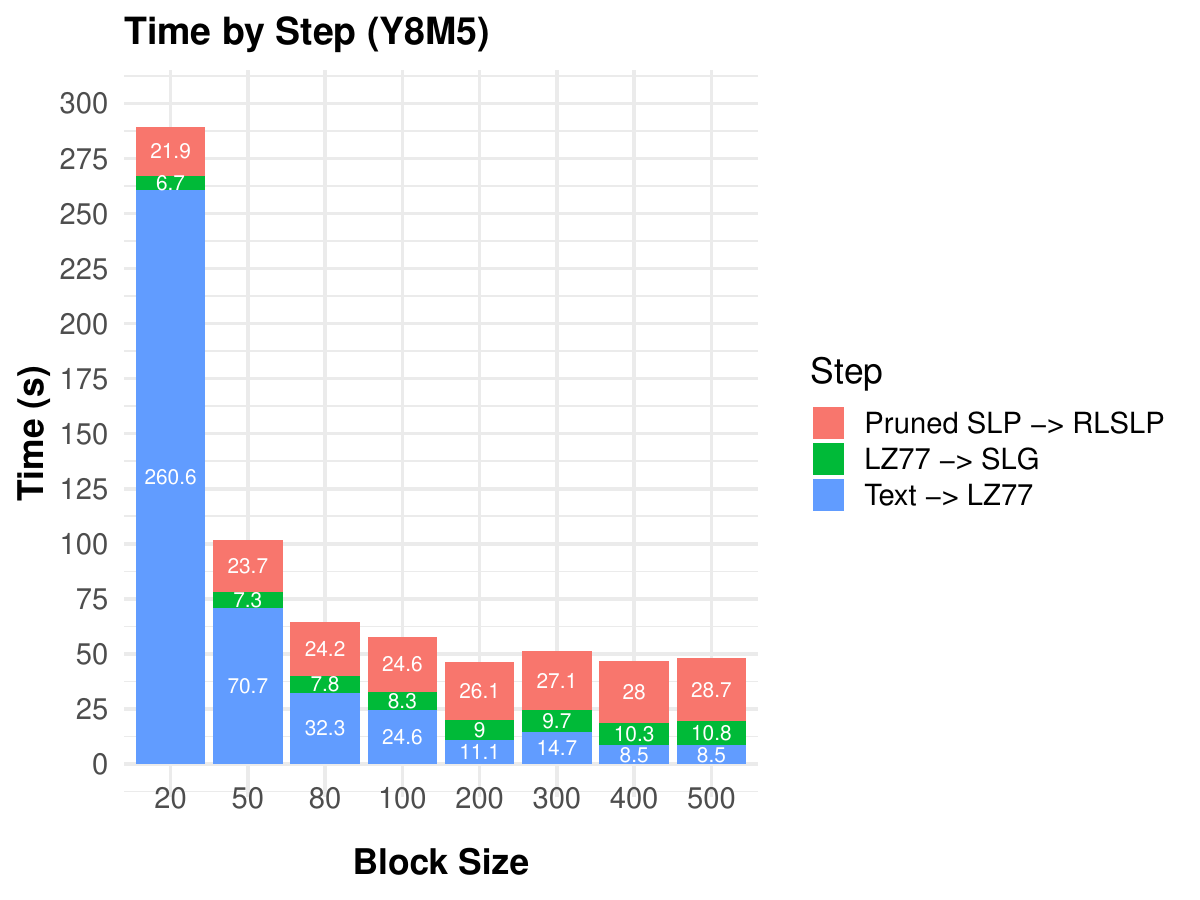}
  \end{minipage}
  \caption{Runtime breakdown of major pipeline components across block
    sizes. Only stages with non-negligible runtime are shown.}\label{fig:timebreakdown}
\end{figure*}

\begin{figure*}[t]
  \begin{minipage}{0.5\textwidth}
    \includegraphics[width=\columnwidth]{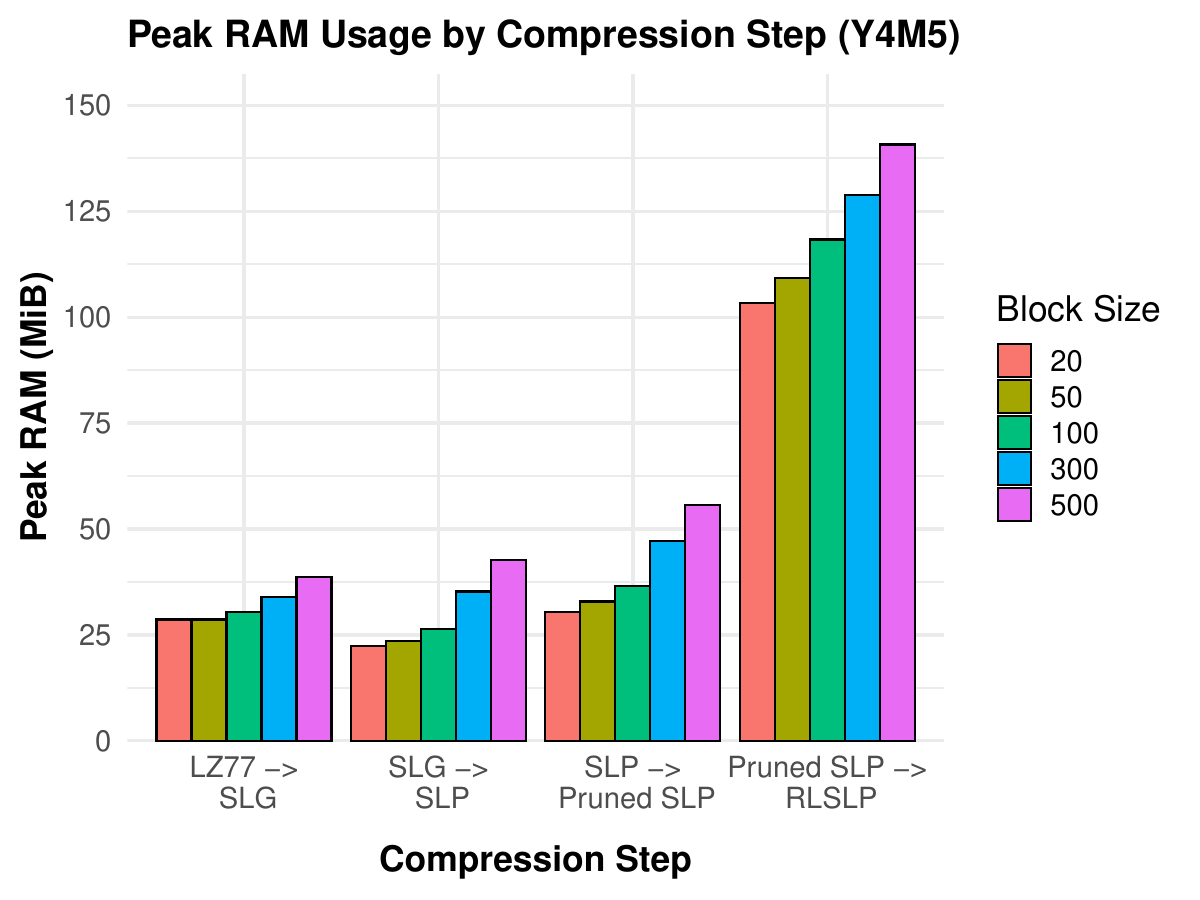}
  \end{minipage}
  \hfill
  \begin{minipage}{0.5\textwidth}
    \includegraphics[width=\columnwidth]{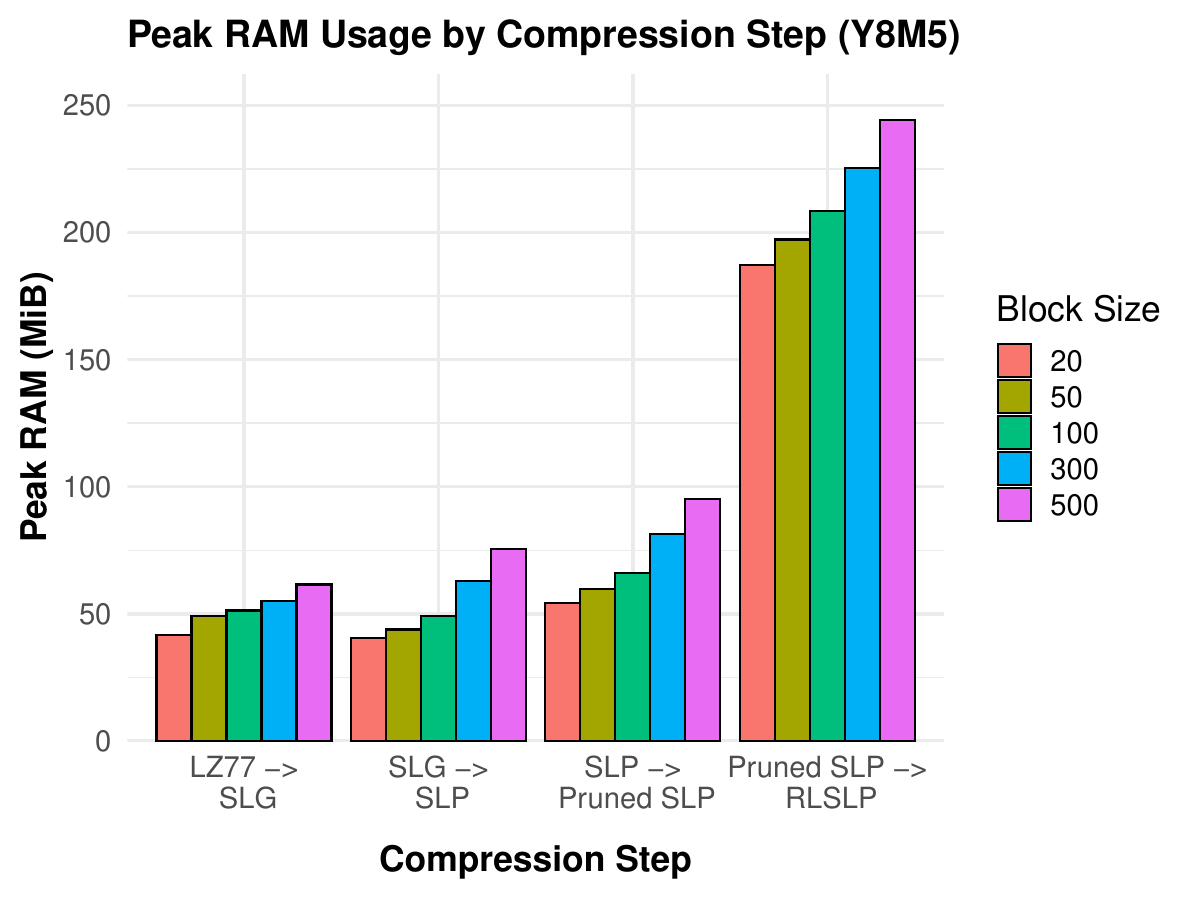}
  \end{minipage}
  \caption{Peak RAM usage of each compressed pipeline component across
    block sizes.}\label{fig:peakram}
\end{figure*}

\subsection{Experiment 3: Breakdown of Runtime in the Compressed Pipeline}\label{sec:experiment3}

This experiment analyzes how runtime is distributed across different
stages of the compressed pipeline. For each block size, we measure the
time spent in five steps: (1) computing the approximate LZ77 from the
input text, (2) converting the approximate LZ77 to an SLG, (3)
expanding the SLG to an SLP, (4) pruning the SLP, and (5) converting
the pruned SLP to a recompression RLSLP. The results are reported on
the Y4M5 and Y8M5 datasets.

Figure~\ref{fig:timebreakdown} shows the breakdown for selected block
sizes. Stages (3) and (4) are omitted from the plot, as their runtimes
are consistently negligible ($<0.5$\,s).

The \texttt{Text} $\rightarrow$ \texttt{LZ77} step dominates runtime
at small block sizes. For Y4M5, at block size~20, it accounts for over 80\,s of
total time. Runtime decreases sharply with increasing block size,
stabilizing below 8\,s beyond block size~100. In contrast, the final
step (\texttt{Pruned SLP} $\rightarrow$ \texttt{RLSLP}) increases with
block size, reaching 13.8\,s at block size~500. The intermediate
\texttt{LZ77} $\rightarrow$ \texttt{SLG} step remains inexpensive
across all settings, with runtime below 7\,s. The results for Y8M5 are
analogous.

These results align with trends from Experiment~2. Small blocks result
in smaller LZ77-like parsings but increase parsing overhead. Larger
blocks reduce front-end cost but shift more work to the back-end
stages.

\subsection{Experiment 4: Peak RAM Usage During the Compressed Pipeline}\label{sec:experiment4}

This experiment measures the peak memory consumption of individual
components in the compressed pipeline, assuming a precomputed
LZ77-like parse as input. This setup models a scenario in which the
LZ77-like factorization is produced externally and the remaining
grammar construction must operate under memory constraints. We report
the peak RAM usage of the four downstream components: \texttt{LZ77}
$\rightarrow$ \texttt{SLG}, \texttt{SLG} $\rightarrow$ \texttt{SLP},
\texttt{SLP} $\rightarrow$ \texttt{Pruned SLP}, and \texttt{Pruned
  SLP} $\rightarrow$ \texttt{RLSLP}. Results are based on the Y4M5 and Y8M5
datasets.

Figure~\ref{fig:peakram} shows peak memory usage across block
sizes. For Y4M5, the final stage (\texttt{Pruned SLP} $\rightarrow$
\texttt{RLSLP}) consistently uses the most memory, increasing from
104\,MiB at block size~20 to 134\,MiB at block size~500. Earlier
stages remain lightweight: \texttt{LZ77} $\rightarrow$ \texttt{SLG},
\texttt{SLG} $\rightarrow$ \texttt{SLP}, and \texttt{SLP}
$\rightarrow$ \texttt{Pruned SLP} all stay well under, or only
slightly above, 50\,MiB. The results on Y8M5 are analogous.

These findings show that memory usage grows moderately with block size
and that the increase is concentrated in the final RLSLP construction
step. This observation is consistent with Experiments~2 and~3: a
larger block size leads to a slightly bigger LZ77-like parse
(Experiment~2), which in turn increases both the runtime of the final
step (Experiment~3) and the peak RAM usage observed here.

\begin{table*}[t]
  \small
  \centering
  \begin{tabular}{@{}llrrr@{}}
  \toprule
  \textbf{Strategy} & \textbf{Dataset} & \textbf{Runtime (s)} & \textbf{Peak RAM (MiB)} & \textbf{RLSLP Productions} \\
  \midrule
  Deterministic & Y4M4 & 87.68 & 699.36 & 4,981,200 \\
  Mixed & Y4M4 & 77.44 & 753.22 & 5,307,093 \\
  Randomized & Y4M4 & 56.20 & 778.78 & 6,080,788 \\
  \midrule
  Deterministic & Y4M5 & 11.37 & 103.36 & 869,055 \\
  Mixed & Y4M5 & 10.36 & 109.33 & 936,193 \\
  Randomized & Y4M5 & 9.00 & 118.31 & 1,073,603 \\
  \bottomrule
  \end{tabular}
  \caption{Comparison of partitioning strategies on Y4M4 and Y4M5 datasets in our compressed pipeline.}
  \label{tab:partitioning}
\end{table*}

\subsection{Experiment 5: Effect of Partitioning Strategy}\label{sec:experiment5}

This experiment examines how different partitioning strategies
(Section~\ref{sec:implementations}) affect the \texttt{Pruned SLP}
$\rightarrow$ \texttt{RLSLP} transformation. We compare deterministic,
randomized, and mixed strategies on the Y4M4 and Y4M5 datasets, using
a block size of~50 in the Bentley--McIlroy step. For each strategy, we record the RLSLP
construction time, peak RAM usage during construction, and the number
of productions in the resulting recompression RLSLP. The results are
summarized in Table~\ref{tab:partitioning}.

The deterministic strategy yields the smallest output RLSLP and the
lowest peak RAM usage, but it also incurs the highest runtime on both
datasets. This is because deterministic partitioning \emph{guarantees}
a constant-factor reduction of the current string, leading to fewer
recompression rounds and a more compact output. However, determining the
partition $\Sigma = \Sigma_{L} \cup \Sigma_{R}$ is computationally
expensive. In contrast, the randomized partition is fast and easy to
compute, yet it reduces the string length less in each PComp step,
requiring more recompression rounds and producing a larger output. The
mixed strategy, which alternates between deterministic and randomized
steps, offers a balance between these two approaches.

\bibliographystyle{siamplain}
\bibliography{paper}

\end{document}